\documentclass{elsart4-1}

\usepackage{amssymb}
\usepackage[english,francais]{babel}
\usepackage{graphicx}
\usepackage{url}

\newtheorem{e-proposition}[theorem]{Proposition}

\newtheorem{e-definition}[theorem]{Definition\rm}

\renewcommand{\theequation}{\arabic{equation}}
\def\og{\leavevmode\raise.3ex\hbox{$\scriptscriptstyle\langle\!\langle$~}}
\def\fg{\leavevmode\raise.3ex\hbox{~$\!\scriptscriptstyle\,\rangle\!\rangle$}}

\begin{document}
% Select a primary header Physics or Astrophysics
% You can place after the header (classification), if you know it.

\centerline{Physics or Astrophysics/Header}
\begin{frontmatter}


% \author{Name\thanksref{label2}}
\selectlanguage{english}
\title{Comets at radio wavelengths}

% If all authors are at the same address, the [label1] can be suppressed

\selectlanguage{english}
\author{Jacques Crovisier},
\ead{jacques.crovisier@obspm.fr}
\author{Dominique Bockel\'ee-Morvan},
\ead{dominique.bockelee@obspm.fr}
\author{Pierre Colom},
\ead{pierre.colom@obspm.fr}
\author{Nicolas Biver}
\ead{nicolas.biver@obspm.fr}

\address{L\'ESIA, Observatoire de Paris, CNRS, UPMC, Universit\'e
Paris-Diderot\\
5 place Jules Janssen, 92195 Meudon, France}
%\address[authorlabel2]{Address2}

% If your know the dates of reception, and acceptation you can put them now;
%    idem the name of the person presenting your article

\medskip
\begin{center}
{\small Received *****; accepted after revision +++++}
\end{center}

\begin{abstract}
Comets are considered as the most primitive objects in the Solar
System.  Their composition provides information on the composition of
the primitive solar nebula, 4.6 Gyr ago.  The radio domain is a
privileged tool to study the composition of cometary ices.
Observations of the OH radical at 18 cm wavelength allow us to measure
the water production rate.  A wealth of molecules (and some of their
isotopologues) coming from the sublimation of ices in the nucleus have
been identified by observations in the millimetre and submillimetre
domains.  We present an historical review on radio observations of
comets, focusing on the results from our group, and including recent
observations with the Nan\c{c}ay radio telescope, the IRAM antennas,
the Odin satellite, the Herschel space observatory, ALMA, and the MIRO
instrument aboard the Rosetta space probe.

{\it To cite this article: J. Crovisier, D. Bockel\'ee-Morvan, P.
Colom, N. Biver, C. R. Physique X (XXXX).}

\vskip 0.5\baselineskip

\selectlanguage{francais}
\noindent{\bf R\'esum\'e}
\vskip 0.5\baselineskip
\noindent
{\bf L'\'etude des com\`etes en ondes radio}

Les com\`etes sont consid\'er\'ees comme les vestiges les mieux
pr\'eserv\'es du Syst\`eme solaire primitif.  Leur composition nous
renseigne sur la composition de la n\'ebuleuse primitive il y a 4,6
milliards d'ann\'ees, fournissant des contraintes sur la formation du
Syst\`eme solaire.  La radioastronomie est un outil privil\' egi\'e
pour l'\'etude des glaces com\'etaires.  Le domaine d\'ecim\'etrique
permet de mesurer la production en eau, par l'observation du radical
OH \`a 18 cm de longueur d'onde.  Le domaine millim\'etrique et
submillim\'etrique permet d'observer de nombreuses mol\'ecules
provenant de la sublimation des glaces du noyau, ainsi que leurs
isotopologues.  Nous pr\'esentons un panorama historique des
d\'ecouvertes sur les com\`etes faites en radioastronomie, mettant
l'accent sur les r\'esultats de notre groupe, et incluant des
observations r\'ecentes avec le radiot\'elescope de Nan\c{c}ay, les
antennes de l'IRAM, le satellite Odin, l'observatoire spatial
Herschel, ALMA, et l'instrument MIRO de la sonde spatiale Rosetta.

{\it Pour citer cet article~: J. Crovisier, D. Bockel\'ee-Morvan, P.
Colom, N. Biver, C. R. Physique X (XXXX).}

%Now keywords/mots-clÈs
 \vskip 0.5\baselineskip
\noindent{\small{\it Mots-cl\'es~:} Com\`etes; Composition chimique;
Formation du Systme solaire; Radioastronomie}}
\end{abstract}
\end{frontmatter}

% now the Version franÁaise abrÈgÈe, if it exists
%\selectlanguage{francais}
%\section*{Version fran\c{c}aise abr\'eg\'ee}
% Text of your Version franÁaise abrÈgÈe here

\selectlanguage{english}
% main text

\section{Introduction}
%=====================

\begin{figure}[h]
\centerline{\includegraphics[height=62mm]{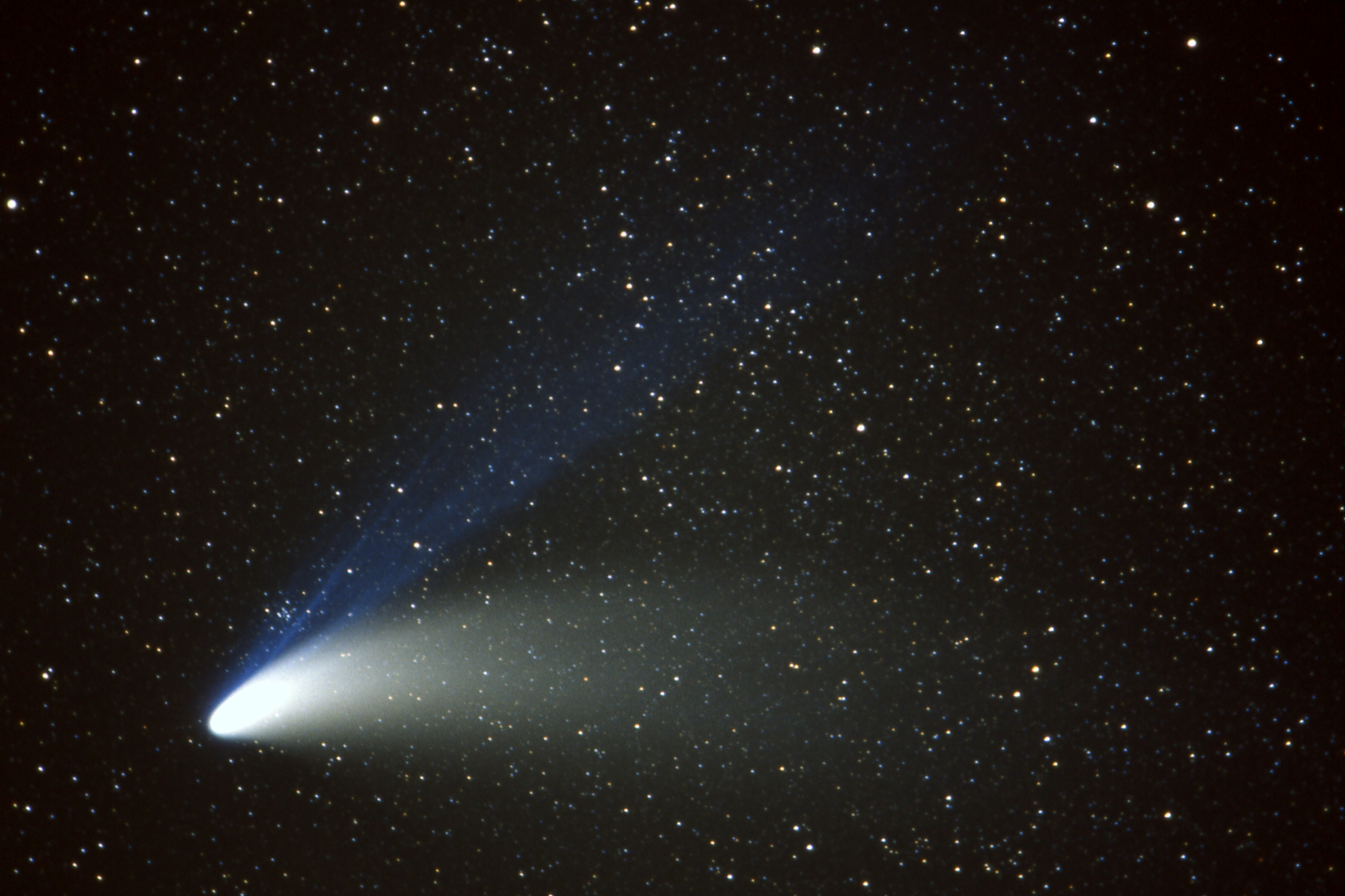}
\includegraphics[height=62mm]{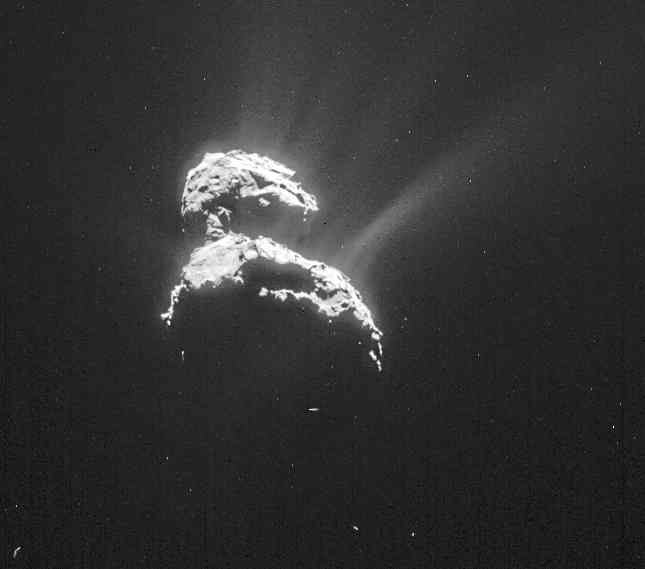}}
\caption{Cometary images.  Left: Comet C/1995 O1 (Hale-Bopp) observed
from the Earth on 6 April 1997 (photo: N. Biver).  The ion and dust
tails are spread here over several $10^{7}$~km (to compare with the
Sun-Earth distance 1~AU = $1.5 \times 10^{8}$ km).  Right: the
nucleus of comet 67P/Churyumov-Gerasimenko observed at a distance of
198~km by the Rosetta space probe on 18 February 2015 (\copyright
ESA/Rosetta/NAVCAM).  The size of the nucleus is about 4~km.  This
image was processed to enhance the dust jets which escape the
nucleus.}
\label{fig-1}
\end{figure}

With a size of a few kilometres, cometary nuclei are practically
unobservable at a distance.  Their direct study requires exploration
by spacecraft.  However, they are icy bodies.  When they approach the
Sun, ices sublimate, releasing gas and dust which form an atmosphere
and tails that may become very spectacular (Fig.~\ref{fig-1}).  As
they contain matter which remained practically intact since the
beginning of the formation of the Solar System, these bodies are
precious testimonies from the past that justify dedicated studies.

Early in the history of radio astronomy, the first attempts to observe
\textit{great comets} C/1956 R1 (Arend-Roland), C/1965 S1
(Ikeya-Seki), C/1969 Y1 (Bennett) and a few others did not yield
probing results.  Indeed, it is not easy to observe a comet at radio
wavelengths.  The signal is weak so that large radio telescopes,
equipped with sensitive receivers, are needed.  The observations
should only be attempted on bright comets, which are rare.  For these
moving objects, one must blindly rely on ephemerides and the telescope
tracking system.  Last but not least, comets are variable,
unpredictable objects, so that an observation can difficultly be
repeated for confirmation.  It is thus not a surprise that the
beginnings of the radio investigation of comets were a succession of
failures, missed opportunities, contradictory results and doubtful
detections that could not be confirmed.

\section{1973: Comet Kohoutek}
%=============================

\begin{figure}[h]
\centerline{\includegraphics[height=60mm]{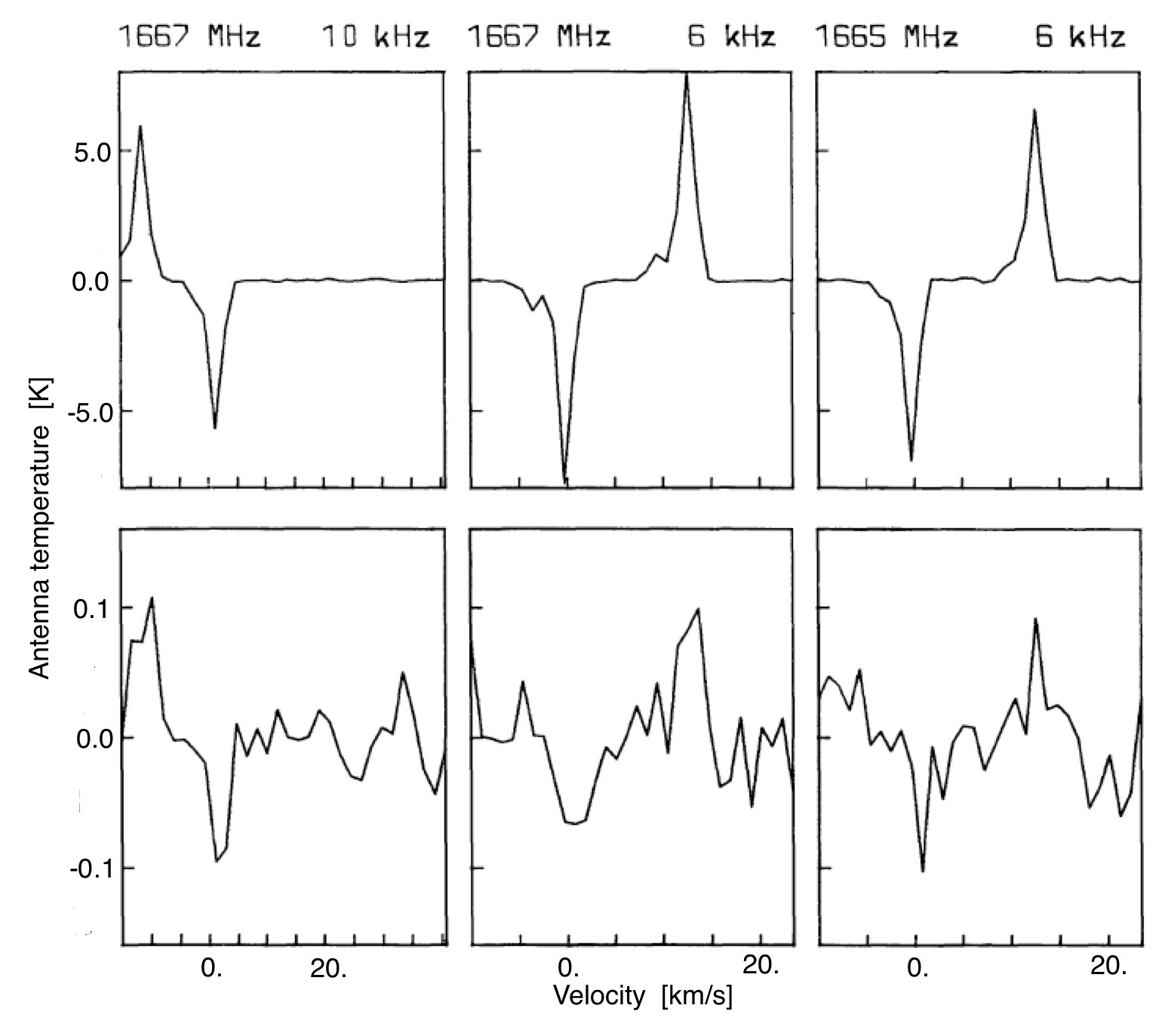}
\includegraphics[height=60mm]{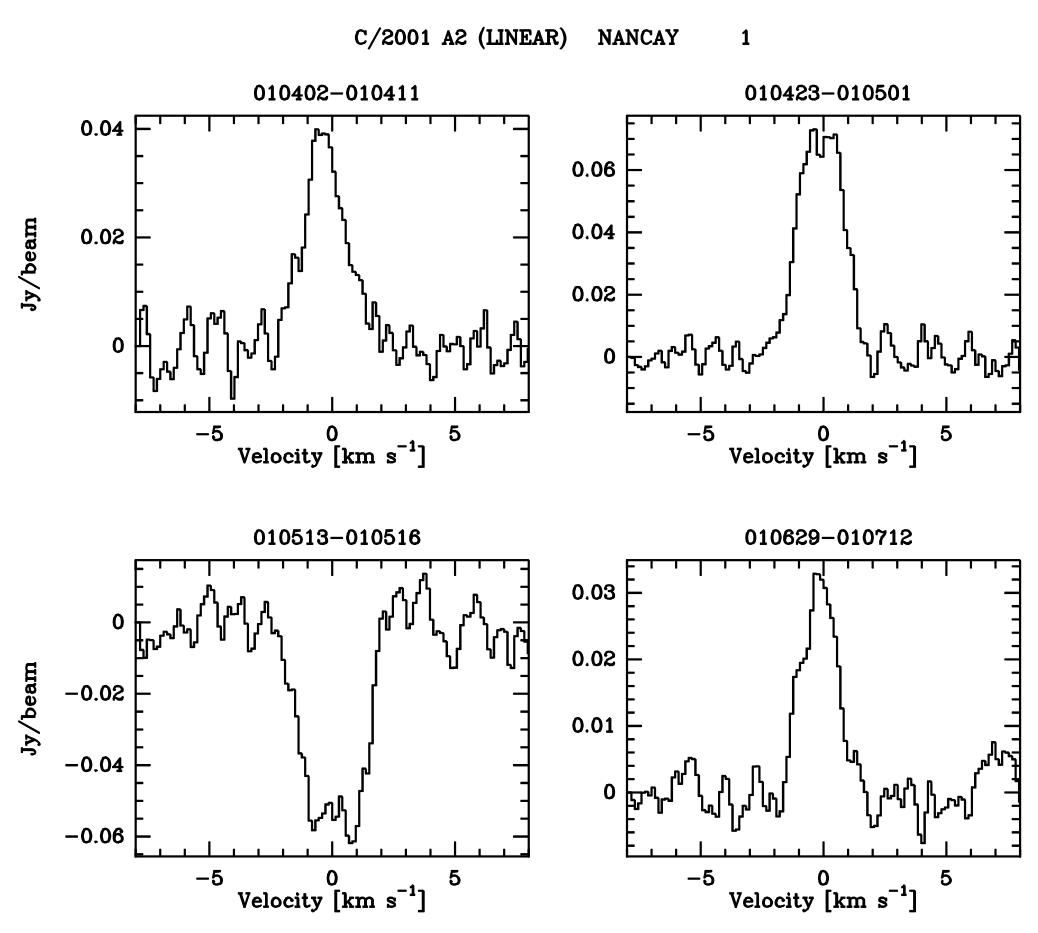}}
\caption{OH spectra from Nan\c{c}ay.  Left: Spectra showing the
detection of the OH lines at 18~cm for the first time in a comet,
C/1973 E1 (Kohoutek), in December 1973 with the Nan\c{c}ay radio
telescope \cite{label1}.  The upper panels show for comparison the OH
lines observed in absorption in the galactic source W12.  The lower
panels show the same lines observed from 1 to 12 December in the
comet.  The observations were performed with frequency switching so
that a positive and negative signal pattern appears.  Right: A
selection of OH spectra obtained at Nan\c{c}ay in comet C/2001 A2
(LINEAR) from April to July 2001 \cite{label3}.  Depending on the
date, the excitation conditions change and the line appears in
emission or in absorption.  In this figure as in most of the spectra
shown in this article, the frequency scale of the x-axis has been
converted to velocities in the frame of the comet following the
Doppler law.  The intensity scales, as is usual in radio astronomy,
are given as antenna temperature or brightness temperature in units of
kelvins, which is the equivalent temperature of a black body in the
Rayleigh-Jean approximation, or flux density in units of janskys (1 Jy
= $10^{-26}$~W~m$^{-2Ñ}$~Hz$^{-1}$).}
\label{fig-2}
\end{figure}

At the end of 1973, the NASA organized a worldwide campaign to observe
comet C/1973 E1 (Kohoutek), in support to its observation aboard
the Skylab orbital station. Following this opportunity, the observation
of the OH radical lines at 18-cm wavelength was attempted at
Nan\c{c}ay.  Their detection was the first detection of a comet at
radio wavelengths (Fig.~\ref{fig-2}, \cite{label1,label2}).

The OH radical is a photodissociation product of water, the major
constituent of cometary ices.  Its observation allows one to determine
the production rate of water, and thus to quantify the activity of the
comet.  This observation of comet Kohoutek was the beginning of a
programme of systematic observations which is still ongoing at the
Nan\c{c}ay radio telescope.  More than 130 comets were thus observed.
The evolution of their activity was followed, in some cases over many
months, to prepare or to complement their observations with other
instruments.

\section{1986: Halley's comet}
%=============================

The historic and mythical Halley's comet (1P/Halley) was in 1986 the target of a
space exploration by no less than five spacecraft. Its observation from the
Earth was also the topic of an international supporting campaign advocated by
the \textit{International Halley Watch}. The radio aspects of this campaign
\cite{label3bis} were coordinated by \' Eric G\'erard (France), William Irvine
and F. Peter Schloerb (United States).

\begin{figure}[h]
\centerline{\includegraphics[height=45mm]{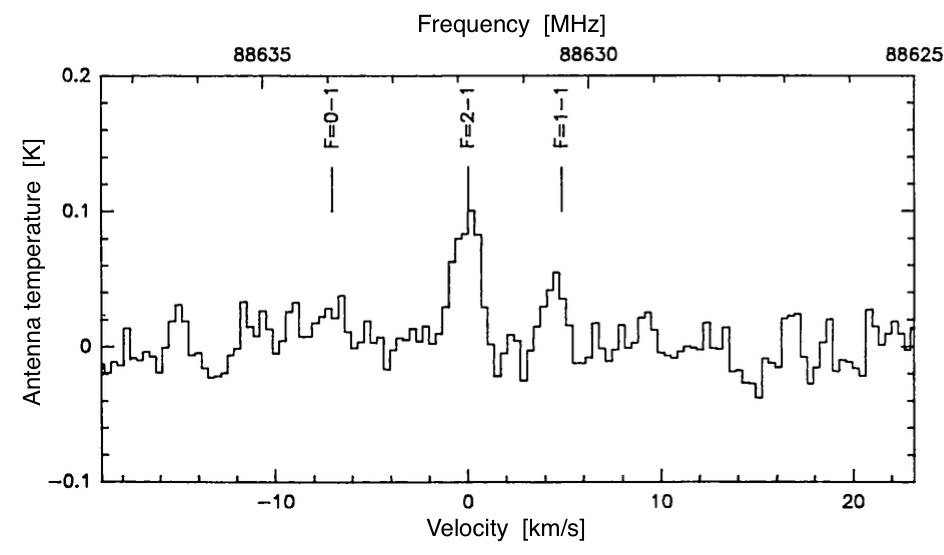}}
\caption{The detection of the three hyperfine components of the $J$
1--0 rotational line at 88.6 GHz of hydrogen cyanide HCN in Halley's
comet with the 30-m antenna of IRAM in November 1985 \cite{label4}.}
\label{fig-3}
\end{figure}

The OH 18-cm lines in Halley's comet have been monitored with the Nan\c{c}ay
radio telescope for a year and a half in 1985--1986 and with several other
telescopes \cite{label3,label3bis}. The same lines were also mapped with the
Very Large Array (VLA) interferometer \cite{label3ter}. Hydrogen cyanide (HCN)
was detected by the newly commissioned 30-m radio telescope of IRAM (Institut de
Radio Astronomie Millim\'etrique) at Pico Veleta (Spain) (Fig.~\ref{fig-3},
\cite{label4}).

This marked the entry of cometary radio astronomy in the world of
molecular radio spectroscopy, already so successful in the study of
interstellar molecules.  A short time after came the
identification, still at IRAM, of methanol (CH$_3$OH) and hydrogen
sulphide (H$_2$S) in comets C/1989 X1 (Austin) and C/1990 K1 (Levy)
\cite{label5}.

\section{1997: Comet Hale-Bopp}
%==============================

\begin{figure}[h]
\centerline{\includegraphics[height=120mm]{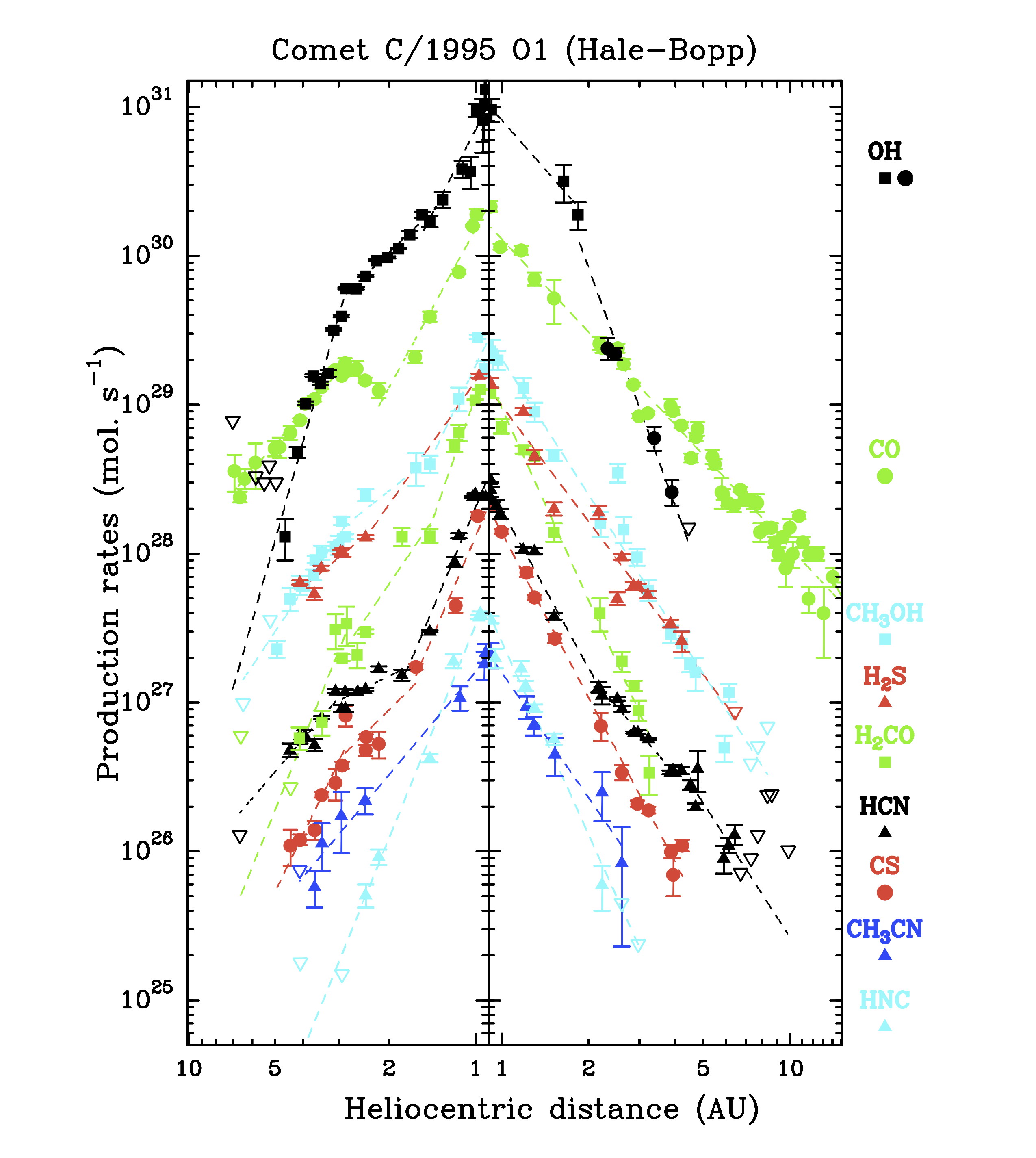}}
\caption{Evolution of the production rates of various molecules in
comet C/1995 O1 (Hale-Bopp) as a function of its distance to the Sun
\cite{label6}.  The left part refers to the pre-perihelion period,
with the comet approaching the Sun, and the right part, to the post
perihelion period, with the comet receding from the Sun.  The data for
the OH radical, which traces the water production, are from Nan\c{c}ay
(squares), with additional post-perihelion values from UV observations
(circles).  The other molecules were observed with the IRAM 30-m
antenna (Spain) or the SEST 15-m telescope (Chile).  Downward pointing
triangles are upper limits.  The sublimation of water appears to be
the dominant process at less than 3~AU from the Sun, whereas carbon
monoxide could be responsible for cometary activity at larger
distances.  The asymmetry of the curves may be due to season effects,
or to thermal evolution of the cometary nucleus.}
\label{fig-4}
\end{figure}

Another exceptional comet was giant comet C/1995 O1 (Hale-Bopp). Its early
discovery, one year and 8 months before its perihelion passage on 1st April
1997, favoured the organization of its observing campaign. It could be detected
at Nan\c{c}ay at the record distance of 4.6~AU from the Sun. Its water
production rate reached at perihelion 300 tons per second, 10 times more than
Halley's comet. The monitoring of the OH production served as a reference for
the other molecular radio observations (Fig.~\ref{fig-4}).

\begin{figure}[h]
\centerline{\includegraphics[height=70mm]{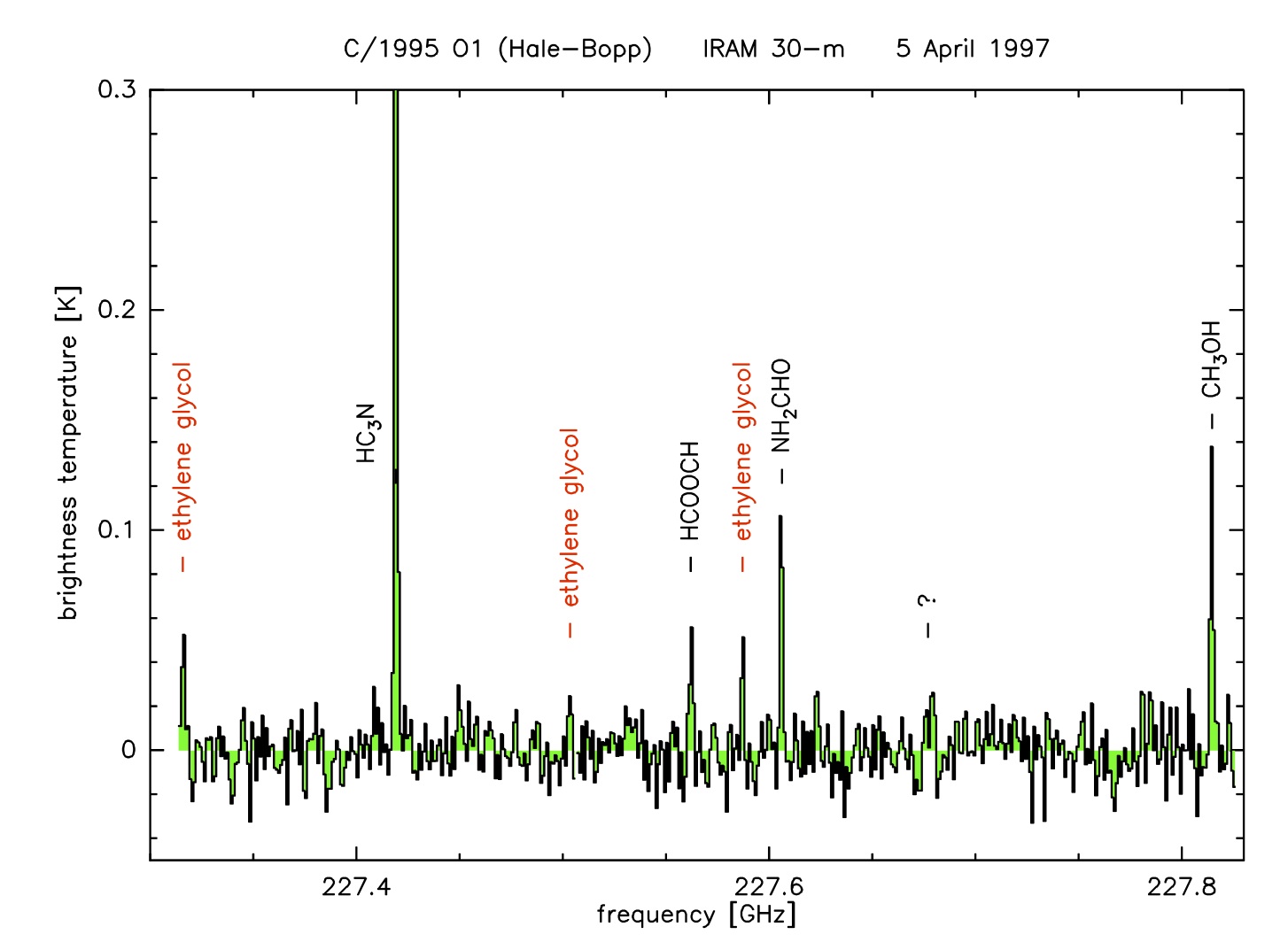}}
\caption{A spectrum of comet C/1995 O1 (Hale-Bopp) observed with the
IRAM 30-m antenna showing the lines of several complex organic
molecules \cite{label8}.  The lines of ethylene glycol (CH$_2$OH)$_2$
were identified seven years after the observation, following the
publication of the laboratory spectrum of this molecule.}
\label{fig-5}
\end{figure}

It is remarkable that among the two dozens of molecular species
released by cometary ices which were then identified in this comet,
two-third were identified by radio spectroscopy (Figs \ref{fig-4},
\ref{fig-5}, \cite{label6,label7,label8}).  In addition to carbon
monoxide, one can note:

\begin{itemize}

\item CHO molecules: H$_2$CO, CH$_3$OH, HCOOH, HCOOCH$_3$,
(CH$_2$OH)$_2$.

\item Nitrogenous molecules: NH$_3$, HCN, HNC, HC$_3$N, CH$_3$CN,
NH$_2$CHO.

\item Sulphuretted molecules: H$_2$S, CS, SO, SO$_2$, OCS, H$_2$CS.

\end{itemize}

In addition, isotopic ratios D/H, $^{12}$C/$^{13}$C, $^{14}$N/$^{15}$N,
$^{32}$S/$^{34}$S were also measured. Most of these detections were made at
millimetric or submillimetric wavelengths with the IRAM radio telescopes (the
30-m antenna in Spain and the interferometer at Plateau-de-Bure), the CSO
(Caltech Submillimeter Observatory 10.5-m antenna) and the JCMT( James Clerk
Maxwell Telescope 15-m antenna) at Hawaii. Ammonia was detected at
centimetric wavelengths with the 100-m radio telescope at Effelsberg (Germany)
\cite{label8bis}. Other molecules which are non-polar, and thus devoid of
rotational lines at radio wavelengths, were detected by infrared spectroscopy:
carbon dioxide and the hydrocarbons CH$_4$, C$_2$H$_2$, C$_2$H$_6$. Another
wealth of identifications of cometary molecules is expected from the in situ
investigations of the mass spectrometers aboard Rosetta.

Radio observations also allow us to probe key physical parameters of
the cometary atmosphere:

\begin{itemize}

\item Its expansion velocity, from the shapes of the lines, taking
advantage of the very good spectral resolution of the radio technique.

\item Its temperature, from the simultaneous observation of the
intensity of several rotational lines of a molecule such as methanol.

\end{itemize}

For comet Hale-Bopp, it was possible to follow the evolution of these
parameters over a large range of heliocentric distances (up to 14~AU;
Fig.~\ref{fig-4}).  The coma expansion velocity was observed to
increase from 0.5 to 1.3 km/s, and the temperature from about 20~K to
130~K, as the comet approached the Sun.  These variations are in
agreement with thermodynamical models of the coma.

\section{Recent observations at IRAM and with ALMA}
%==================================================

The versatility of the spectrometers now equipping the IRAM 30-m radio
telescope is precious for cometary observations which are
time-critical and which cannot be easily repeated.  With an
instantaneous frequency coverage of $2 \times 8$~GHz, spectral surveys
are now easily feasible, allowing simultaneous observations of many
molecules and serendipitous detections.  In recent comets C/2012 F6
(Lemmon) and C/2013 R1 (Lovejoy), this possibility allowed us to
retrieve some molecules which up to now were only spotted in comet
Hale-Bopp \cite{label14}.  And in comet C/2014 Q2 (Lovejoy), two new
molecules --- ethanol (C$_2$H$_5$OH) and glycolaldehyde (CH$_2$OHCHO)
--- were detected in January 2015 \cite{label14bis}.

\begin{figure}[h]
\centerline{\includegraphics[height=70mm]{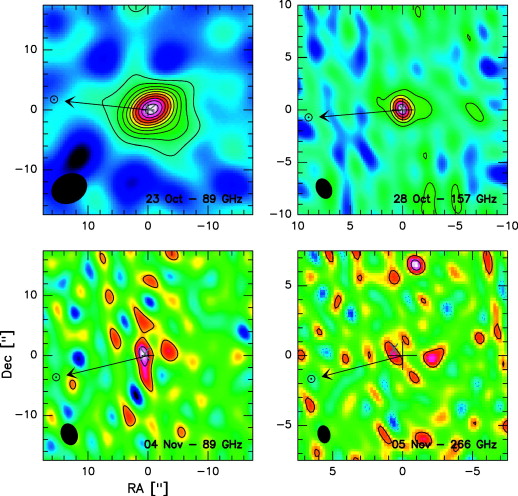}}
\caption{Maps of continuum emission at 1, 1.5 and 3~mm wavelength of
comet 103P/Hartley 2 observed with the IRAM interferometer at
Plateau-de-Bure in October-November 2010 \cite{label15}, when this
comet passed at only 0.12 AU ($1.8 \times 10^{7}$~km) from the Earth.
These maps show the distribution of large-size dust particles in the
coma and the continuum emission of the nucleus.  The arrows show the
direction of the Sun.}
\label{fig-6}
\end{figure}

Continuum emission as well as molecular line emissions can be mapped by
radio interferometers.  The cometary continuum is dominated by the
thermal emission of dust in the coma, the signal from the nucleus
being very difficult to detect with Earth-based observations, except for comets coming close to the Earth. Continuum emission from dust particles is only significant at wavelengths smaller than the particle size.
Thus its observation at millimetric wavelengths characterizes large-size
particles. This is complementary to visible and infrared observations
which are sensitive to particles of very small sizes.  

Mapping
molecular lines allows us to know the distribution of these molecules
within the coma and to probe coma chemistry.  The origin of the gas
molecules may thus be traced: they are either coming directly from the
nucleus, or progressively injected in the coma following chemical
reactions or the sublimation of icy grains.  Asymmetries in their
distribution may be due to \textit{jets} originating from
\textit{active regions} of the nucleus' surface.

\begin{figure}[h]
\centerline{\includegraphics[height=90mm]{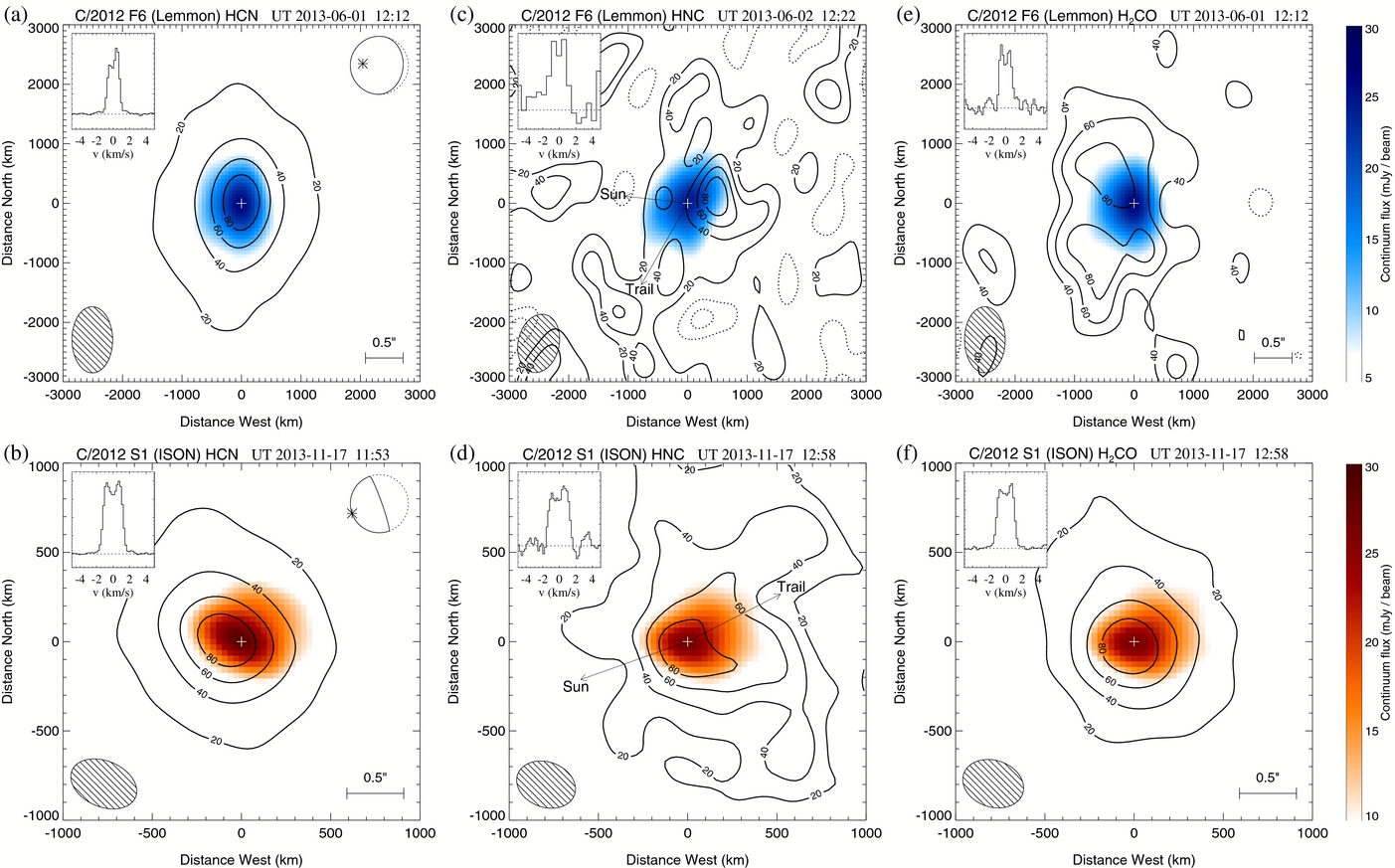}}
\caption{From left to right, maps of the emission lines of hydrogen
cyanide HCN, its isomer HNC and formaldehyde H$_2$CO observed with
ALMA in comets C/2012 F6 (Lemmon) (top) and C/2012 S1 (ISON) (bottom)
\cite{label16}.  Spectra of the individual lines are shown in the
upper left inserts.  These maps show that molecules such as HNC and
H$_2$CO are progressively released in the coma from still ill-known
sources, whereas HCN seems to come directly from the nucleus ices.}
\label{fig-7}
\end{figure}

The IRAM interferometer at Plateau-de-Bure consists presently of six
15-m antenna which are to be extended to 12 antenna with the NOEMA
(NOrthern Extended Millimeter Array) project.  Its first significant
results were obtained on comet Hale-Bopp in 1997 \cite{label15A,label15B}, then at the occasion of the outburst of comet 17P/Holmes \cite{label15C} and of the passage close to the Earth of comet 107P/Hartley~2
(Fig.~\ref{fig-6},  \cite{label15}).  The Atacama Large
Millimeter/submillimeter Array (ALMA) in Chile, with its 50 12-m
antennas, is more sensitive.  Its first cometary observations were
performed in 2013 on comets C/2012 F6 (Lemmon) and C/2012 S1 (ISON).
(Fig.~\ref{fig-7}, \cite{label16}).

%\newpage

\section{Space radio telescopes: SWAS, Odin, Herschel}
%=====================================================

Although water is the prime constituent of cometary ices, its observation is
difficult. Its rotational lines, which are in the submillimeter range, are not
observable from the ground due to the opacity of the Earth's atmosphere. Their
detection in comets is relatively recent.

\begin{figure}[h]
\centerline{\includegraphics[height=50mm]{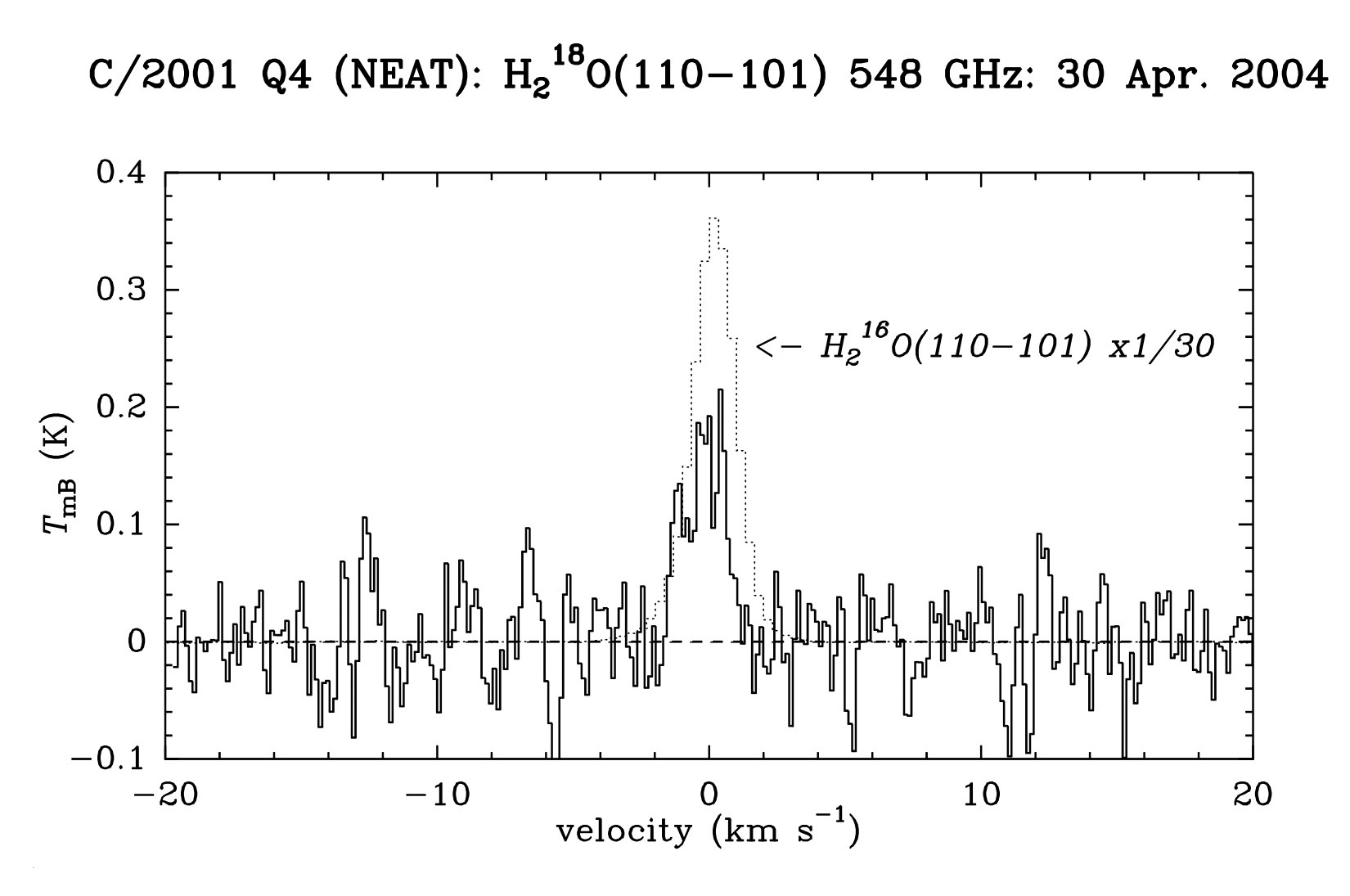}}
\caption{The water lines of the water isotopologues H$_2^{16}$O and
H$_2^{18}$O observed by the Odin satellite in comet C/2001 Q4 (NEAT)
\cite{label10}.}
\label{fig-8}
\end{figure}

The fundamental submillimetric line of water at 557~GHz (the $1_{10}$--$1_{01}$
transition at 0.5~mm wavelength) was finally observed from two satellites
dedicated to the study of this line, the Submillimeter Wave Astronomy Satellite
(SWAS, launched by the United States in 1998), and Odin (launched by the Swedish
space agency in 2001, with a French participation consisting in the delivery of
an acousto-optic spectrometer). Odin observed water in a dozen of comets. It
also observed NH$_3$ and H$_2^{18}$O, measuring for the first time by remote sensing the
$^{18}$O/$^{16}$O isotopic ratio in a comet. Odin is still operational and
observed comet C/2014 Q2 (Lovejoy) in February 2015. (Fig.~\ref{fig-8},
\cite{label9,label10})

\begin{figure}[h]
\centerline{\includegraphics[height=38mm]{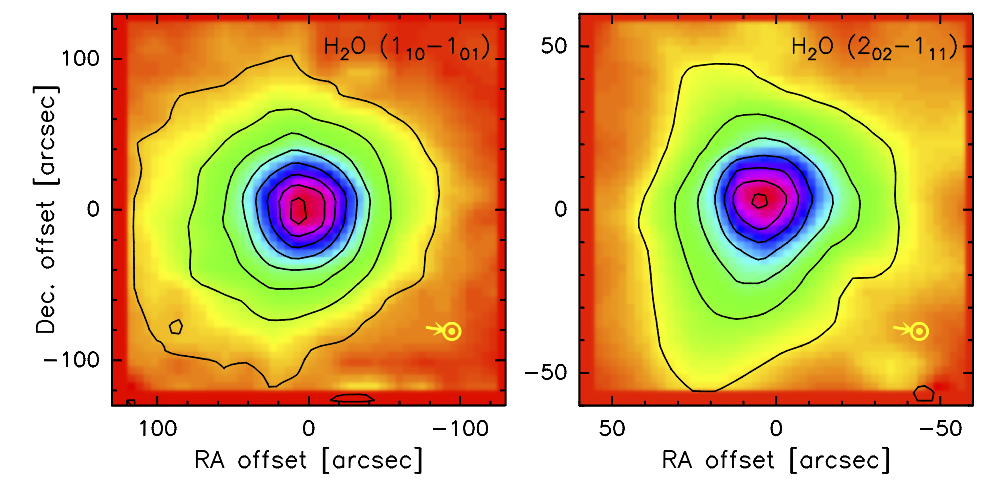}
\includegraphics[height=38mm]{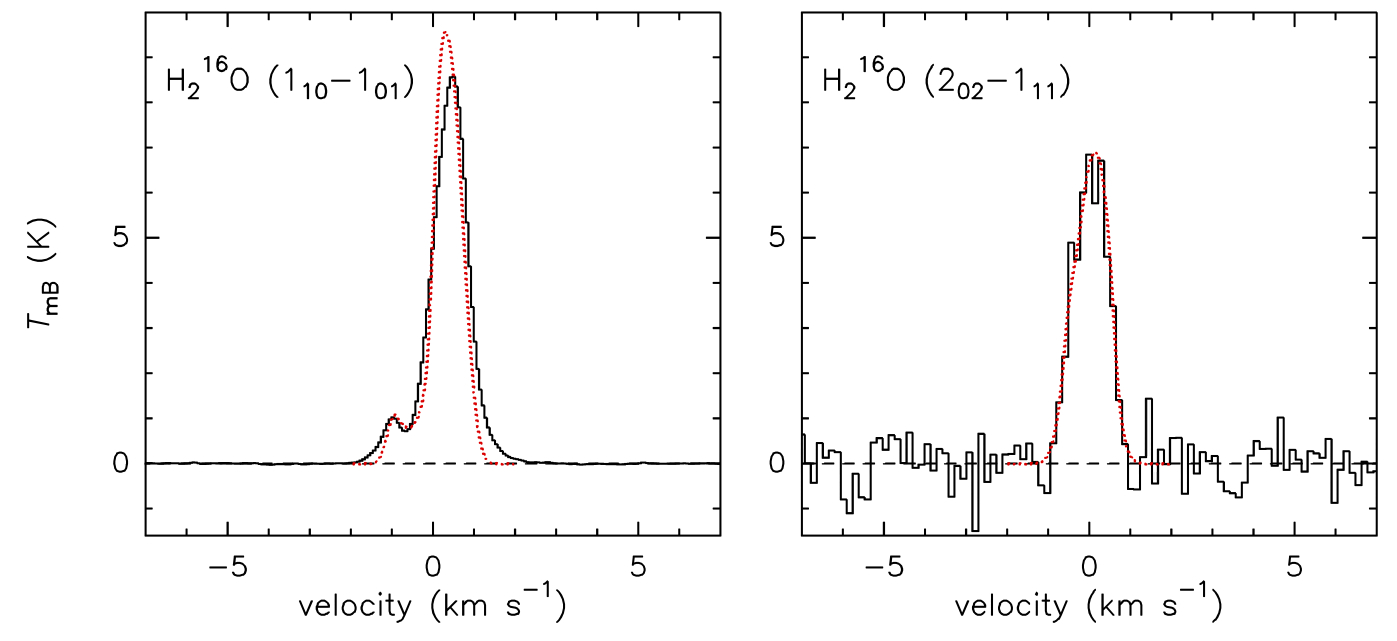}}
\caption{Maps and profiles of water lines in comet C/2009 P1 (Garradd)
observed with the Herschel space observatory \cite{label13}.  The 
$1_{10}$--$1_{01}$ line at 557 GHz is not symmetric: this very strong line
is saturated, showing an indentation at negative velocities caused by
self-absorption from the foreground external layers of the cometary
atmosphere.}
\label{fig-9}
\end{figure}

The next step was the Herschel space observatory, operating in the
submillimeter and far-infrared domains, launched by the European Space
Agency.  With its 3-m diameter mirror and its liquid-helium cooled
instruments (including HIFI, the Heterodyne Instrument for the Far
Infrared), it was much more sensitive than its predecessors.  From
2009 to 2013, it could study and map several lines of water in a
dozen of comets, some of them being weak, distant comets.  In three of
them, it measured the D/H isotopic ratio.  (Figs~\ref{fig-9}, \ref{fig-12},
\cite{label11,label12,label13})

\section{Close observation of a comet with MIRO, the radio telescope
aboard Rosetta}
%===================================================================

\begin{figure}[h]
\centerline{\includegraphics[height=60mm]{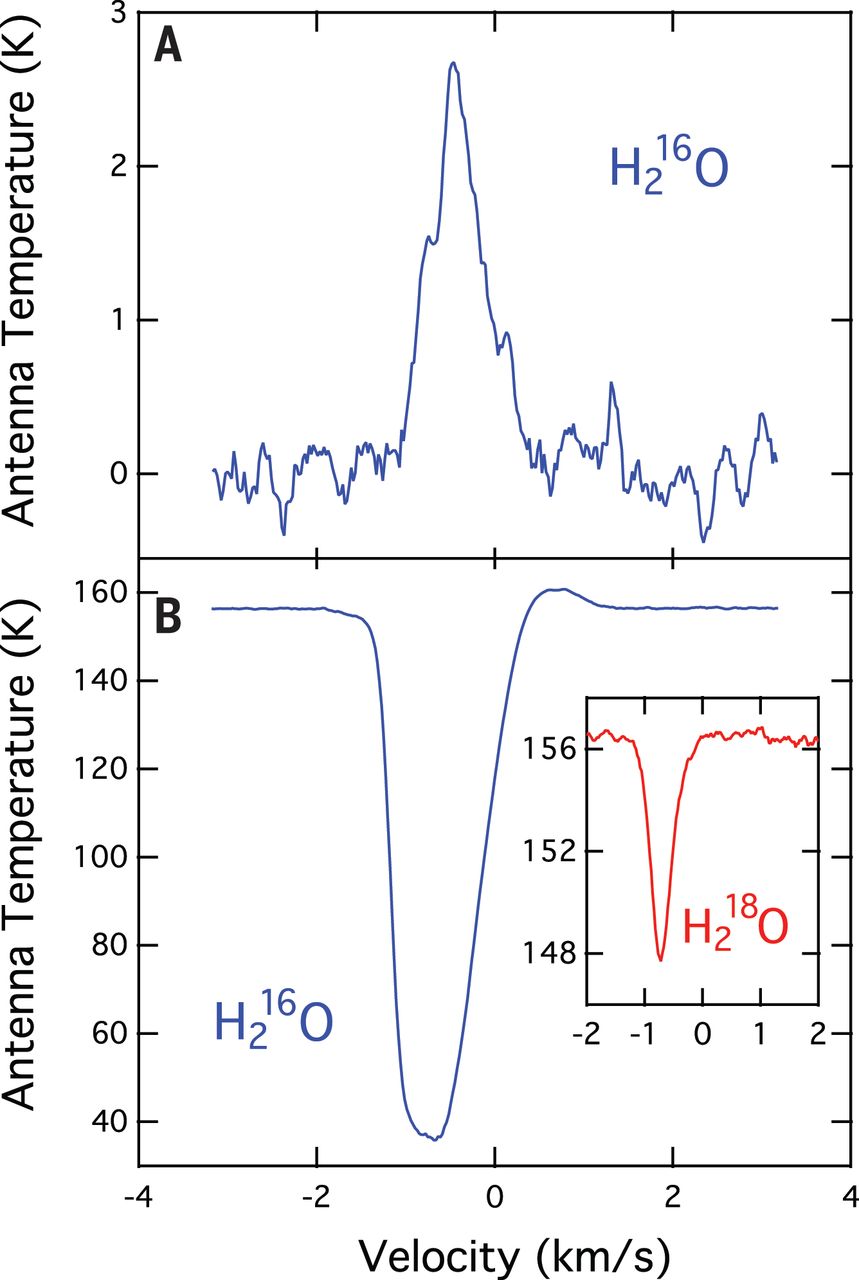}}
\caption{An example of the water lines observed at 0.5~mm wavelength in comet
67P/Churyumov-Gerasimenko with MIRO on Rosetta \cite{label18}. Top: observation
on 23 June 2014 at a distance of 128\,000~km. Bottom: observation on 19 August
2014 at a distance of 81~km; the lines are then seen in absorption against the
continuum of the nucleus.}
\label{fig-11}
\end{figure}

\begin{figure}[h]
\centerline{\includegraphics[height=50mm]{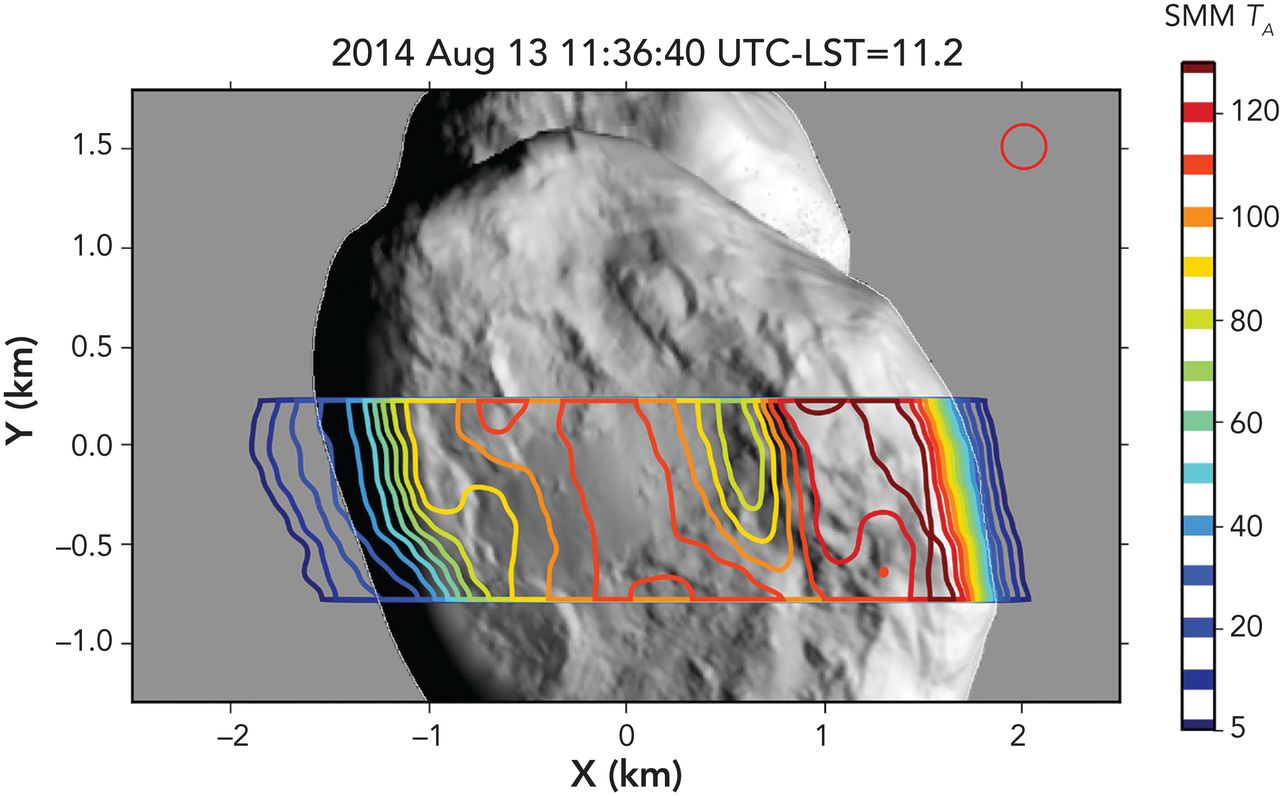}}
\caption{A partial map of the temperature of the nucleus of comet
67P/Churyumov-Gerasimenko, from the submillimetric continuum observations of
MIRO on Rosetta, superimposed on the model of its shape deduced from imaging in
the visible \cite{label18}. It can be seen on the left side that MIRO observes
the night side for which visible imaging provides no information. The
temperatures vary, following solar insolation, from about 30~K (night side) to
130~K. The comet was then at 3,5 AU from the Sun.}
\label{fig-10}
\end{figure}

The Rosetta space probe, launched by the European Space Agency in 2004, is
exploring comet 67P/Churyumov-Gerasimenko, orbiting its nucleus at distances
which, at some moments, were as close as 8~km. It is equipped with a dozen of
instruments, one of them being MIRO (Microwave Instrument for the Rosetta
Orbiter), a radio telescope with an antenna of only 30~cm diameter. In the
vicinity of the comet, this modest size is enough to study a selection of lines
of water (several isotopic species), methanol, ammonia and carbon monoxide with
a spectrometer operating around 0.5~mm wavelength. MIRO is also equipped with
two continuum channels at 0.5 and 1.6~mm wavelength, dedicated to the
observation of the thermal emission of the nucleus. (Figs \ref{fig-10},
\ref{fig-11}, \cite{label17,label18})

Rosetta and its instruments are to monitor the evolution of the comet
until at the end of August 2016 and to follow its activity, which
climaxed at its perihelion in August 2015.

\section{Other studies and perspective}
%======================================

This rapid survey, mainly focussed on the results from our group, is
far from exhausting all aspects of the study of comets at radio
wavelengths.  One should also mention:

\begin{itemize}

\item Continuum observations with bolometers on single dish telescopes, sensitive to dust grains in the coma \cite{label18bis}.

\item Radar studies of cometary nuclei, which determine the size and
shape of these objects when they come within the range of large
Earth-based antennas \cite{label19}.

\item The study of the plasma environment of comets, where the solar
wind and the cometary coma interact \cite{label19bis}.

\item The velocimetry of cometary space probes, still the only way to
evaluate precisely the mass of cometary nuclei.  Compared with the
nucleus dimensions obtained by imaging, the nucleus density can then
be obtained.  Rosetta thus determined a density of 533~kg~m$^{-3}$
for 67P/Churyumov-Gerasimenko \cite{label20}.

\item The tomography of the nucleus of 67P/Churyumov-Gerasimenko
performed by the CONSERT instrument, a bi-static radar operating
between the Rosetta orbiter and its lander Philae \cite{label21}.

\item Laboratory measurements of cometary matter analogues, for the
interpretation of radar and radiometric observations of cometary
nuclei \cite{label22}.

\end{itemize}

The space exploration of comets, due to its complexity and its high
cost, will be limited for a long time to a small number of targets,
which precludes statistical investigations.  These latter can only be
done from Earth-based systematic observations.  The observations of
several tens of comets, and especially spectroscopic radio
observations, have shown that the relative productions of water,
carbon monoxide, methanol and many other molecular species may highly
differ from one object to the other.  This suggests that comets have
different chemical compositions, but we do not know yet how these
differences are related to the history of the formation and evolution of
these bodies.

\begin{figure}[h]
\centerline{\includegraphics[height=70mm]{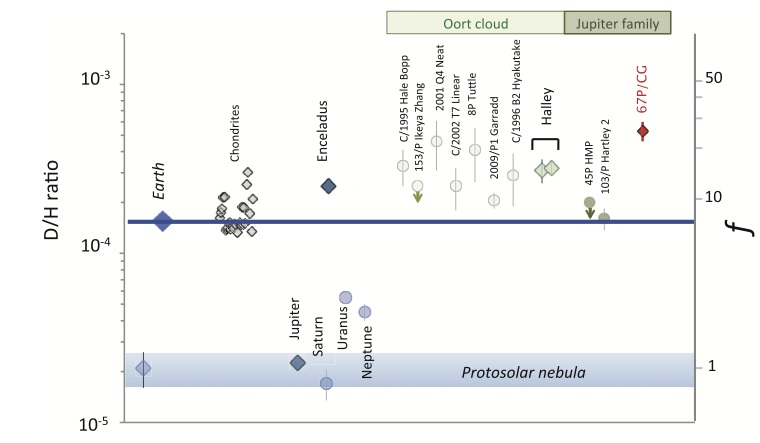}}
\caption{The D/H ratio observed in cometary water and in other Solar
System bodies.  Most of the cometary data are from radio observations.
The cometary D/H ratio ranges from one (103P/Hartley 2 observed
with Herschel) to three times (67P/Churyumov-Gerasimenko observed with
the ROSINA mass spectrometer of Rosetta) its value in Earth oceans
\cite{label12,label13,label23}.}
\label{fig-12}
\end{figure}

The D/H isotopic ratio of cometary water, compared to that of Earth
oceans, is customarily used to test the hypothesis of a cometary
origin for terrestrial water.  This ratio is presently only known for
a small number of comets (from radio observations for several of
them).  These D/H values range from one to three times the terrestrial
value, which seems to rule out an origin from only comets
(Fig.~\ref{fig-12}, \cite{label12,label13,label23}).  In the future, a
better knowledge of this statistic will help us to better know to
which extent the fall of comets, asteroids and other small bodies on
Earth contributed to terrestrial water.  Further constraints on the
origin of comets and cometary material are also provided from the D/H
and other isotopic ratios which are measured in molecules other than
water.

% etc, etc

% \appendix

% \section{}
% \label{}

% The Acknowledgements are also a un-numbered section
%\section*{Acknowledgements}
% Acknowledgements text here

%Figures: \ref{fig-1}, \ref{fig-2}, \ref{fig-3}, \ref{fig-4}, \ref{fig-5}, \ref{fig-6}, \ref{fig-7}, \ref{fig-8}, \ref{fig-9}, \ref{fig-10}, \ref{fig-11}, \ref{fig-12}, \ref{fig-13}.

\end{document}